\documentclass[a4paper,fleqn]{cas-dc}

\AtBeginEnvironment{table}{\small}
\usepackage[numbers]{natbib}
\usepackage{graphicx} 
\usepackage{caption}
\usepackage{tikz}
\usetikzlibrary{positioning, shapes}
\usetikzlibrary{trees}
\usepackage{multirow}
\usepackage{pgfplots}
\pgfplotsset{compat=1.18}
\usepackage{longtable}
\usepackage{enumitem}
\usepackage{booktabs}
\usepackage{tabularx}
\usepackage{ragged2e}
\usepackage{pdflscape}
\usepackage{geometry}
\usepackage{enumitem}
\usepackage{array}
\usepackage{xcolor}
\usepackage{fontawesome}
\usepackage[most]{tcolorbox}

\def\tsc#1{\csdef{#1}{\textsc{\lowercase{#1}}\xspace}}
\tsc{WGM}
\tsc{QE}
\tsc{EP}
\tsc{PMS}
\tsc{BEC}
\tsc{DE}

\begin{document}
\let\WriteBookmarks\relax
\def\floatpagepagefraction{1}
\def\textpagefraction{.001}

\shorttitle{Age-related app reviews}

\shortauthors{S. Nirmani et~al.}

\title [mode = title]{Age Matters: Analyzing Age-Related Discussions in App Reviews}

\author[1]{Shashiwadana Nirmani}[]
\cormark[1]
\ead{s.dona@research.deakin.edu.au}
\affiliation[1]{organization={Deakin University},
    addressline={Burwood}, 
    city={VIC},
    country={Australia}}

\author[1]{Garima Sharma}[]
\ead{garimasharma2007@gmail.com}

\author[1]{Hourieh Khalajzadeh}[]
\ead{hourieh.khalajzadeh@deakin.edu.au}

\author[2]{Mojtaba Shahin}[]
\ead{mojtaba.shahin@rmit.edu.au}

\affiliation[2]{organization={RMIT
University},
    addressline={Melbourne}, 
    city={VIC},
    country={Australia}}

\cortext[cor1]{Corresponding author}

\begin{abstract}
In recent years, mobile applications have become indispensable tools for managing various aspects of life. From enhancing productivity to providing personalized entertainment, mobile apps have revolutionized people's daily routines. Despite this rapid growth and popularity, gaps remain in how these apps address the needs of users from different age groups. Users of varying ages face distinct challenges when interacting with mobile apps, from younger users dealing with inappropriate content to older users having difficulty with usability due to age-related vision and cognition impairments. Although there have been initiatives to create age-inclusive apps, a limited understanding of user perspectives on age-related issues may hinder developers from recognizing specific challenges and implementing effective solutions.
In this study, we explore age discussions in app reviews to gain insights into how mobile apps should cater to users across different age groups. We manually curated a dataset of 4,163 app reviews from the Google Play Store and identified 1,429 age-related reviews and 2,734 non-age-related reviews. We employed eight machine learning, deep learning, and large language models to automatically detect age discussions, with RoBERTa performing the best, achieving a precision of 92.46\%. Additionally, a qualitative analysis of the 1,429 age-related reviews uncovers six dominant themes reflecting user concerns: \textit{Age Appropriateness of Content, Language and Recommendations, Age Verification and Access Barriers, Usability and Accessibility Across Ages, Privacy and Safety Concerns, Interactions and Relationships} and \textit{Recommendations and Feature Requests}. 
\textcolor{black}{ Our findings reveal that users frequently encounter inappropriate content for children, struggle with strict or error-prone age verification systems, and emphasize the need for age-friendly accessibility and safety features. To address these issues, we offer actionable recommendations for app developers, including implementing flexible or gradient-based age restrictions, prioritizing safety features in kids' apps, and strengthening parental controls.}
\end{abstract}




\maketitle

\section{Introduction}

Mobile applications (apps) have become an integral part of people's lives in areas ranging from shopping and banking to education and entertainment. As of April 2025, the App Store and Google Play Store have 2 million apps each \cite{GooglePlay}\cite{AppleAppStore}. With this ever-growing number of mobile apps and their users, the diversity of users continues to grow. Mobile app users vary widely in terms of different factors such as age, ethnicity,  gender, level of education, and socioeconomic status \cite{ramos2021considerations}\cite{khalajzadeh2022supporting}. This rapid growth and diversification in app usage highlight significant gaps in how apps cater to users from different age groups. In this work, we specifically focus on how age is represented and discussed in mobile app reviews, aiming to uncover age-related concerns and experiences expressed by users.

As mobile app usage among children and seniors increases, the need for age-appropriate apps becomes critical. Individuals of different ages face distinct problems when using software and may also respond differently to the same software \cite{grundy2021impact}. Younger individuals who are more familiar with technology tend to have different expectations of apps compared to older adults \cite{williams2013considerations}. Children may be exposed to inappropriate content and form interactions that could compromise their safety. They also have cognitive and motor skill limitations that impact how they interact with digital interfaces \cite{yadav2022designing, clemente2024digital}. Designing for children requires consideration of their limited reading skills, short attention spans, and tendency to exploratory behavior, which can lead to unintentional actions like accidental in-app purchases or navigating to unsafe content \cite{amaefule2023fostering, booton2023impact,meyer2019advertising,clemente2024digital}. 

 Ageing brings physical changes, especially in sensory functions, affecting the ability to interact with technical systems. For example, vision deteriorates with age, including reduced visual acuity, colour perception, contrast sensitivity, and increased sensitivity to glare \cite{wirtz2009age}. Adult users may struggle with complex interfaces or lack of accessibility features \cite{esafety_inappropriate_content,vic_inappropriate_content,chen2013children,zhao2023mobile,diaz2014accessibility}. When designing software interfaces for older adults, these changes should be carefully considered by selecting colours, contrast, fonts, sizes and designs of interface elements to ensure better usability \cite{wirtz2009age}. Overlooking the age group of end users may result in software that is overly complex or unclear, has inferior interfaces and workflow, and is insufficiently engaging or enjoyable \cite{grundy2021impact}.  
 
 It is important to determine how demographic factors like age influence the apps' user experience, user preferences, and challenges. Many apps fail to consider these age-specific requirements, making improvements necessary for broader aspects such as user experience and safety. Age discussions in mobile apps can range from inadequate parental controls to poor readability and navigation for the elderly, all of which contribute to a less inclusive digital environment. Despite their importance, age-related factors remain overlooked in both research and the design of commercial products \cite{sacar2024designing}. 

 Several studies have explored age-related concerns in software, including usability issues, design considerations for children, and accessibility for older users \cite{jim2021improving,wirtz2009age,eisenberg1995creating,sim2006all,diaz2014accessibility}. These works emphasize that age-friendly design involves more than just visual elements, highlighting the importance of interaction design and clear feedback. However, none have specifically examined how age is discussed in app reviews

The Google Play Store and Apple App Store allow users to provide feedback on the apps using reviews. They can provide their feedback through star ratings, comments, or both. These review data provide insights into the user's perception of the app \cite{khalid2014mobile}. Further, these app reviews can highlight critical issues developers may not have anticipated, offering a window into the practical challenges different age groups encounter when interacting with apps. Comprehending such review feedback will enable mobile app developers to address users' concerns, ultimately increasing the ratings of the apps \cite{khalid2014mobile}. 

\textcolor{black}{However, age-related concerns are currently buried within millions of app reviews and are not systematically surfaced to developers. Manual analysis of the reviews is impractical on a large scale. Therefore, automatically identifying age discussions is crucial for understanding and responding to the diverse needs of users at different stages of life. By detecting these discussions automatically, our goal is to support developers, product teams, and platform moderators in recognising age-specific usability, safety, and content issues early, ultimately improving user experience and inclusiveness.}

This work explores age discussions in mobile app reviews. We used the dataset collected by \textcolor{black}{Shahin et al.} ~\cite{shahin2023study}, which includes 7 million app reviews from 70 popular Android apps in the Google Play Store. We then manually constructed and annotated a dataset of 4,163  reviews, comprising 1,429 age-related reviews and 2,734 non-age-related reviews. We employed \textcolor{black}{eight} Machine Learning (ML), Deep Learning (DL), and Large Language Models (LLM)  to distinguish age-related app reviews from non-age-related app reviews. The RoBERTa classifier achieved the best performance. We achieved precision, recall, F1-score, and accuracy of 92.70\%, 92.39\%, 92.45\%, and 92.39\%, respectively. In the second stage of the study, we manually analyzed 1,429 age-related user reviews and identified that age-related app review discussions revolve around  \textit{Age Appropriateness of Content, Language and Recommendations}, \textit{Age Verification and Access Barriers, Usability and Accessibility Across Ages, Privacy and Safety Concerns, Interactions and Relationships} and \textit{Recommendations and Feature Requests} topics.

The key contributions of this work are: 
\begin{itemize}
    \item We are the first to delve into age discussions in mobile app reviews. 
    \item We develop ML, DL and LLM models to detect age discussions in app reviews automatically.
    \item We develop a comprehensive understanding of different types of age discussions in mobile app reviews.
    \item We provide recommendations for the development and research of apps that are inclusive of age-related aspects. 
\end{itemize}

The rest of the paper is organized as follows: Section \ref{related-work} summarizes the related work, and Section \ref{method} introduces our research methodology. In Section \ref{rq1} and Section \ref{rq2}, we elaborate on the approaches and results of our two research questions. We discuss our reflections on the results in Section \ref{discussion}. In Section \ref{threats}, we report the threats to validity. We conclude the paper in Section \ref{conclusion}.

\section{Related Work} \label{related-work}

\subsection{Age in Software Systems}

Several researchers have studied users' age-specific issues and the development of age-adaptive software systems.  Jim et al. \cite{jim2021improving} investigated how age-related factors can be better integrated into software modelling frameworks. They proposed an extension to wireframe-based designs to address age considerations within the modelling process.  Moreover, the impact of age on designing software interfaces has been explored in several studies \cite{gossen2015knowledge,williams2013considerations,awada2017multimodal,al2012touch,jochmann2010children}. Wirtz et al. \cite{wirtz2009age} examined age-related usability issues that older adults (55+) face by contrasting their navigation performance with youngsters to identify age-specific and exclusive challenges. On the other hand, Gossen et al. \cite{gossen2015knowledge}  designed an age-adaptable search interface with touch-based interactions to meet the evolving needs of various age groups. Williams et al. \cite{williams2013considerations} studied age-related challenges faced by older users when interacting with computing devices. They emphasized the importance of designing interfaces that allow for user-specific customization.   

Researchers have explored the role of age in different mobile app domains. For example,
Natarajan et al. \cite{natarajan2018moderating} studied the factors influencing mobile shopping adoption concerning consumer age and device type. The results revealed that young consumers (\(\leq 35\) years) prefer simple lag-free interfaces. Older consumers (\textgreater{ 35 years}) were influenced by perceived usefulness, satisfaction, and risk on usage intention. They recommended adding useful features and ensuring a risk-free experience with trust-inducing elements like privacy, security, and vendor information to encourage mobile shopping among older users. Gurtner et al. \cite{Gurtner2014} studied designing mobile business apps for different ages. They specifically investigated whether users from different generations have similar or different reasons for adopting new mobile business apps. Young users (up to 25 years) who have grown up in a mobile and digital environment, value convenience and access to services at any time and place rather than ease of use. Middle-aged (26-49 years) users show a significant influence of enjoyment on the perceived usefulness of apps. For mature users (50 plus), ease of use is crucial due to their less experience with new technology and possibly declining absorptive capacity.

Several studies have explored the development of mobile apps for elderly users, identifying various challenges. The identified challenges mainly revolve around design inadequacies, usability limitations and age-specific needs. The majority of fitness and medical apps  are primarily designed for younger users. Hence, they fail to cater to the specific needs of senior citizens \cite{gao2010mobile}. Elderly users visually struggle with small, unclear  text, icons, images, and poor colour contrasts \cite{elguera2019elderly,Darvishy2017}. Usability issues such as fast, repetitive movements, complex tasks requiring many steps, unnecessary animations, and the lack of clear instructions hinder app usage \cite{elguera2019elderly}. Further, clarity, simplicity, and security are crucial for meeting the needs of older adults \cite{Darvishy2017}.

Several researchers have studied the development of children-centric mobile apps. Heikkinen et al. \cite{heikkinen2022designing} studied designing mobile apps for children and provided lessons, development tips and recommendations for developers. Masooda and Thigambaram \cite{masood2015usability} studied the importance of usability in the UI design of mobile educational apps for children aged 4-5. They found that usability is a multi-stage problem-solving process, examining how children interact with the app and their initial impressions and learning experiences. They also provided design guidelines for developers to improve UIs for children's educational apps. Saadiah et al. \cite{latiff2019user} examined the key design elements needed for creating UIs for children and provided a set of guidelines. These guidelines cover eight elements: Navigation, Image and Icon, Text, Content, Audio, Colour, Feedback, and Input/Output Support. Each element includes essential characteristics to help designers create effective UIs for children's mobile learning apps.

\subsection{Human-Centric Discussions in App Reviews}

Several attempts have been made to explore human-centric discussions in app reviews in the last few years (e.g., \cite{arony2023inclusiveness,shahin2023study,khalajzadeh2022supporting,nasab2024study,fazzini2022characterizing,tjikhoeri2024best,obie2021first,nema2022analyzing}). Arony et al. \cite{arony2023inclusiveness} investigated age as one of the demographic factors in app reviews while creating a taxonomy for inclusiveness from discussions on Twitter, Reddit, and Google Play Store.  They identified age as a common sub-category in demography, noting that users often complain about age-related policy issues. The results revealed that age-inclusive issues are often associated with the type of service. They also used five DL classifiers to distinguish inclusiveness-related discussions from other types of discussions automatically. However, unlike our work, they did not focus on developing a dedicated method for identifying age-related reviews, categorizing age discussions, or determining the underlying reasons.

Shahin et al. \cite{shahin2023study} examined discussions about gender from the users' viewpoint in app reviews and offered insights and suggestions for developing apps that are gender inclusive. Khalajzadeh et al. \cite{khalajzadeh2022supporting} explored the human-centric issues that end-users report in Google App Store reviews. They established a taxonomy that organizes these concerns into three primary categories: “App Usage,” “User Reaction”, and “Inclusiveness”. Additionally, they created ML and DL classifiers to automatically identify and classify these user-centric issues found in app reviews. Fairness issues highlighted in reviews of AI-based mobile applications were investigated by Nasab et al. \cite{nasab2024study}. They identified six types of fairness concerns related to AI-based applications. Further, they explored the underlying reasons that app owners provided to explain these fairness issues. Additionally, they developed and tested ML and DL models to identify and differentiate fairness-related reviews from other types of reviews. 

Fazzini et al. \cite{fazzini2022characterizing} conducted an empirical study examining the influence of human factors on software usage, specifically analyzing app reviews for COVID-19 contact tracing apps. Their findings revealed that these apps frequently have difficulty meeting the needs of users across various age groups, which was evident in their ratings. Half of the age-related reviews were negative, while only about 28\% were positive. The study highlights that apps intended for broad community use, such as COVID-19 apps, must consider users of all ages. Many users reporting age-related issues shared their experiences or suggested new features or improvements.

Despite existing literature, there has been no research focused on automatically identifying mobile app reviews that include age discussions or exploring the types of these discussions and their possible underlying causes from the users' perspective. This study aims to address this gap by analyzing user reviews of 70 popular Android apps available on the Google Play Store.

\section{Research Methodology} \label{method}
This work aims to analyze age discussions in mobile app reviews. To achieve this, we proposed the following two research questions (RQs): \\
\textbf{RQ1:} Can age-related reviews be detected automatically and accurately?

\noindent \textbf{RQ2:} What types of age discussions exist in mobile app reviews?

In the scope of the above-mentioned RQs, we define \textbf{age discussions} as \emph{discussions associated with different age groups, including any concern, requirement, or issue from a mobile app user perspective}. This interpretation of age discussions is used further in this work.

\subsection{Dataset}
\label{sec_dataset}
\begin{figure*}
    \centering
    \includegraphics[width=0.95\linewidth]{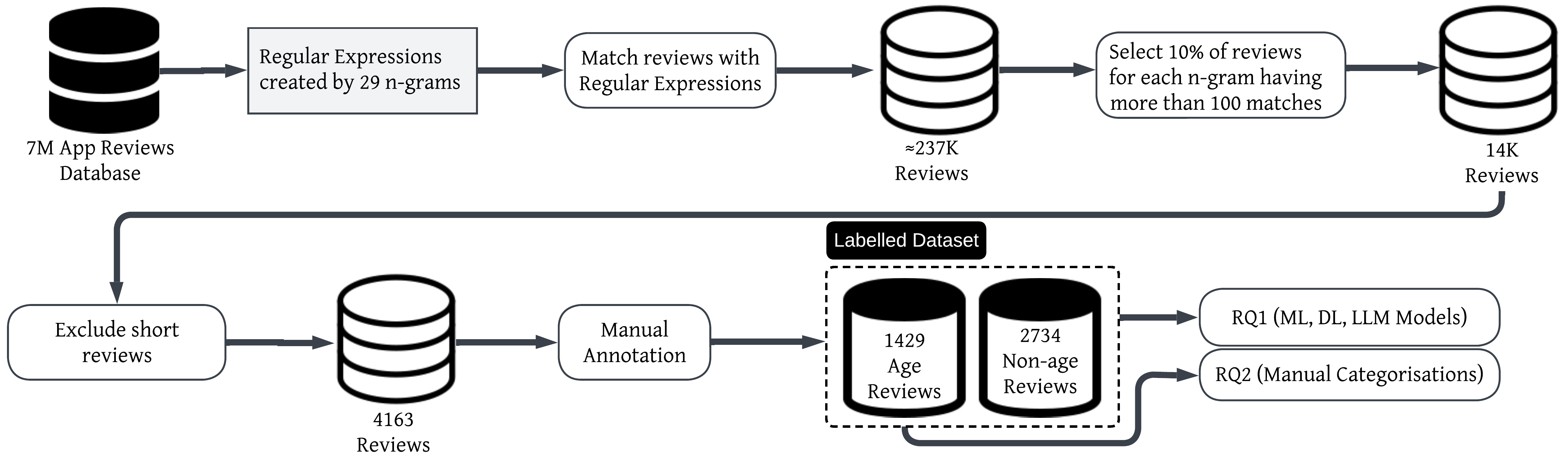}
    \caption{The dataset construction process for RQ1 and RQ2}
    \label{fig:steps}
\end{figure*}

To verify our RQs, we constructed a dataset using the following steps. Figure \ref{fig:steps} shows the dataset construction process elaborated in the following sections.

\subsubsection{Data Collection} 
To the best of our knowledge, there is no publicly available dataset with age-related app reviews. Initially, we explored Android mobile apps targeting people of different age groups (e.g., Elder Launcher: UI for Seniors\footnote{ \url{https://play.google.com/store/apps/details?id=xyz.arjunsinh.elderlauncher}}, Lumosity: Brain Training\footnote{ \url{https://play.google.com/store/apps/details?id=com.lumoslabs.lumosity}}, Emoha - Support for Seniors\footnote{ \url{https://play.google.com/store/apps/details?id=com.emoha}}). However, reviews under these apps were very limited in size and focused only on kids or older people. Thus, these reviews did not adequately represent age-related issues faced by people in different age groups.
Our work uses the 7 million (7M) user reviews collected by~\cite{shahin2023study} from 70 popular Android apps in the Google Play Store. 
These 7M user reviews were used in~\cite{shahin2023study} to identify the gender-related user reviews and to categorize gender discussions in mobile app reviews. In this paper, we used the original 7M user reviews collected in this process without any further annotation done at the original work. 

\subsubsection{Identifying age-related reviews}
After creating a large dataset of mobile app user reviews, the next step was to identify the \emph{potential age-related reviews} to align with our defined RQs. 
As it was challenging to annotate the 7M user reviews directly to find the age-related reviews, inspired by~\cite{nema2022analyzing,shahin2023study,nasab2024study}, we created an age-related taxonomy to simplify this process.

Based on the general understanding of the English language and relevant age-related studies~\cite{portenhauser2021mobile,gomez2023design,elguera2019elderly,borjesson2015designing}, we curated an initial list of 29 distinct n-grams for the age-related taxonomy, presented in Table \ref{ngram}. The list includes terms referring to specific age groups (e.g., children, teenager,  middle aged, senior), familial roles linked with age (e.g., grandparents, parents, grandchildren), and broader concepts associated with ageing and development (e.g., lifespan, life stage, retirement). Our aim was to include a mix of biological, social, and generational terms that are likely to surface in user reviews, ensuring that both direct and nuanced references to age are effectively identified through this taxonomy. \textcolor{black}{To avoid missing synonymous expressions, we iteratively refined the n-gram list by reviewing retrieved samples and incorporating additional terms identified across multiple rounds of pilot inspection.}

These n-grams were then used to create regular expressions, commonly used to search for and match specific text patterns in large datasets quickly and accurately. We created regular expressions to capture all variations of the n-grams. For example, the regular expression for `kid' would also match `kids'. Some examples of regular expressions we used are:
{
\small
\begin{verbatim}
\\b(?:aging|aging|AGING)\\b
\\b(?:adolescent|adolescent|ADOLESCENT)\\b
\end{verbatim}
}

The regular expressions were matched with the user reviews of the dataset (i.e., 7M reviews), which resulted in $\approx$237K app reviews (i.e., potential age-related reviews). The n-grams with at least 100 user reviews were shortlisted for the next steps.

\begin{table*}[]
\caption{List of n-grams}
\label{ngram}
\begin{tabular}{|l|l|l|l|l|}
\hline
adolescent    & age          & aging        & child         & childhood experiences \\ \hline
children      & chronological & elderly     & generation    & grandchild            \\ \hline
grandchildren & grandparents & kid          & life cycle    & life stage            \\ \hline
lifespan      & longevity    & mature       & middle aged   & old                   \\ \hline
parents       & retired      & retirement   & senior        & teenage years         \\ \hline
teenager      & underage     & young        & youth         &                       \\ \hline
\end{tabular}%
\end{table*}

\subsubsection{Data Annotation}
After selecting the potential age-related user reviews from the regular expressions, the next step was to annotate the reviews manually. From $\approx$237K potential age-related reviews, we selected 14K reviews for manual annotation. To achieve this, we selected 10\% of the user reviews corresponding to each n-gram having more than 100 matched user reviews. This ensured that we had a significant proportion of reviews from each n-gram while maintaining an adequate dataset size. We then excluded short reviews containing one, two, or three words from these 14K user reviews, which resulted in 4,163 user reviews.  

4,163 reviews were independently annotated by two coders (i.e., two authors). The two coders were asked to read these reviews and categorize them as age reviews or non-age reviews. In total, 331 disagreements between two coders were found, which were resolved by the third coder (i.e., author). Cohen’s kappa was calculated to assess the level of agreement between the first two coders \cite{mchugh2012interrater}. The resulting kappa value was 0.83, indicating a strong level of agreement. All the coders are experienced in human-computer interaction and social aspects of software engineering. Also, discussions were carried out at different stages between the three coders to have a shared understanding of the age-related concepts and maximize the annotation quality. At the end of this process, 1,429 age reviews and 2,734 non-age reviews were found.

\section{Automatic Detection of Age-related Reviews (RQ1)} \label{rq1}
This section discusses RQ1. We explore different learning models to classify whether a user review is age-related or not. 

\subsection{Dataset}
We used the dataset introduced in Section \ref{sec_dataset}, containing 4,163 user reviews. Of these reviews, we labelled 1,429 as age-related, while the rest were non-age discussions. These annotated user reviews are further used to train our classifiers.

\subsection{Models}
We used a combination of ML, DL, and LLM models to classify the app reviews as age-related. These models are selected based on recent state-of-the-art software engineering and natural language processing methods \cite{nema2022analyzing, hadi2023evaluating, nasab2024study}. Many advanced methods for language classification have proved to be better than the traditional ML techniques. Hence, more focus is placed on the recent DL and LLM models, along with a few ML-based classifiers. 
\subsubsection{Traditional ML Models}
For ML models, we used the commonly used \emph{Universal Sentence Encoder (USE)} features from the user reviews. USE is a transformer-based model specially designed for transfer learning to different Natural Language Processing (NLP) tasks~\cite{cer2018universal}. These models are trained on sentences, short phrases, and short paragraphs from different data sources. This study uses the USE models on user review sentences to extract 512-dimensional embeddings.
We used two popular ML-based classifiers, Support Vector Machine (SVM) \cite{cortes1995support} and eXtreme Gradient Boosting (XGBoost) \cite{chen2016xgboost}, to classify the user reviews. SVM classifies the data by finding a hyperplane to separate the data points of different classes. It uses different kernel functions to handle different types of data. This study uses the polynomial kernel function while using the SVM for classification. On the other hand, XGBoost is an ensemble-based gradient-boosting algorithm that combines multiple weak learners to create a robust classifier.

\subsubsection{Pre-trained Language Models} The following pre-trained Language models were used in this study. 

\textbf{Bidirectional Encoder Representations from Transformers (BERT)}: BERT~\cite{devlin2018bert} is a bidirectional transformer model designed to pre-train bidirectional representations of the text by including context from both the left and right sides. Hence, the BERT models can be fine-tuned for various NLP tasks. This study uses the BERT tokenizer and \emph{bert-base-uncased} model from the Hugging Face library\footnote{ \url{https://huggingface.co/docs/transformers/en/model_doc/bert}} for the classification.  

\textbf{Robustly Optimized BERT Pre-training Approach (RoBERTa)}: RoBERTa~\cite{liu2019roberta} is an extension of BERT models created after modifying the pre-training method in the original BERT models. These improved settings allow RoBERTa models to achieve better performance in the downstream tasks than the BERT models. 
In this study, we used the RoBERTa tokenizer\footnote{\url{https://huggingface.co/docs/transformers/en/model_doc/roberta}} to convert the user reviews into embeddings using byte-level Byte-Pair-Encoding. Further, the \emph{roberta-base} model is used to train the model for the final classification.  

\textbf{DistilBERT: A distilled version of BERT}: DistilBERT~\cite{sanh2019distilbert} is a fast and lighter version of BERT developed by Hugging Face. DistilBERT was trained by the distillation of pre-trained BERT, which is done by training a model to predict the same probabilities as BERT. This helps DistilBERT to achieve comparable results with BERT with 40\% fewer parameters and improved speed.   
In this study, similar to BERT, we used DistilBertTokenizer\footnote{\url{https://huggingface.co/docs/transformers/en/model_doc/distilbert}} to convert the user reviews into useful embeddings. Further, \emph{distilbert-base-uncased} model is used to train the model to classify the user reviews.

\subsubsection{Large Language Models (LLMs)}
The following LLMs were used to read and analyze the user reviews and provide the label as age-related or not, due to their popularity and strong performance in natural language processing tasks \cite{chang2024survey, pasupuleti2024popular}. 

\textbf{Generative Pre-training Transformer (GPT):} GPT is a unidirectional transformer-based model trained with casual language modelling, which allows it to achieve high performance in predicting the next token in a sequence \cite{radford2018improving}. These models are highly efficient in text generation and are highly used in question-answering apps. Specifically, the \emph{gpt-4.1} model was used from the OpenAI library\footnote{ \url{https://platform.openai.com/docs/models/gpt-4.1}} for these experiments.

\textbf{Gemini:} Gemini is a transformer-based multimodal LLM developed by Google \cite{team2023gemini}. It is designed to handle and integrate multiple input types, including text and images, enabling more complex reasoning and interaction. Trained using a combination of supervised fine-tuning and reinforcement learning, Gemini demonstrates strong performance in tasks such as code generation, text completion, and question answering. We employed \emph{Gemini 2.0 Flash} model\footnote{\url{https://ai.google.dev/gemini-api/docs/models\#gemini-2.0-flash}}
for our experiments.

\textbf{Large Language Model Meta AI (LLaMA):} LLaMA is a family of transformer-based language models developed by Meta, optimized for research and open access \cite{touvron2023llama}. These models are trained using standard autoregressive language modeling, aiming to deliver high performance across a wide range of natural language understanding and generation tasks \cite{llama2025}. We utilized \emph{Llama-4-Scout-17B-16E-Instruct}\footnote{\url{https://huggingface.co/meta-llama/Llama-4-Scout-17B-16E-Instruct}} in this study.

\begin{figure}[h!]
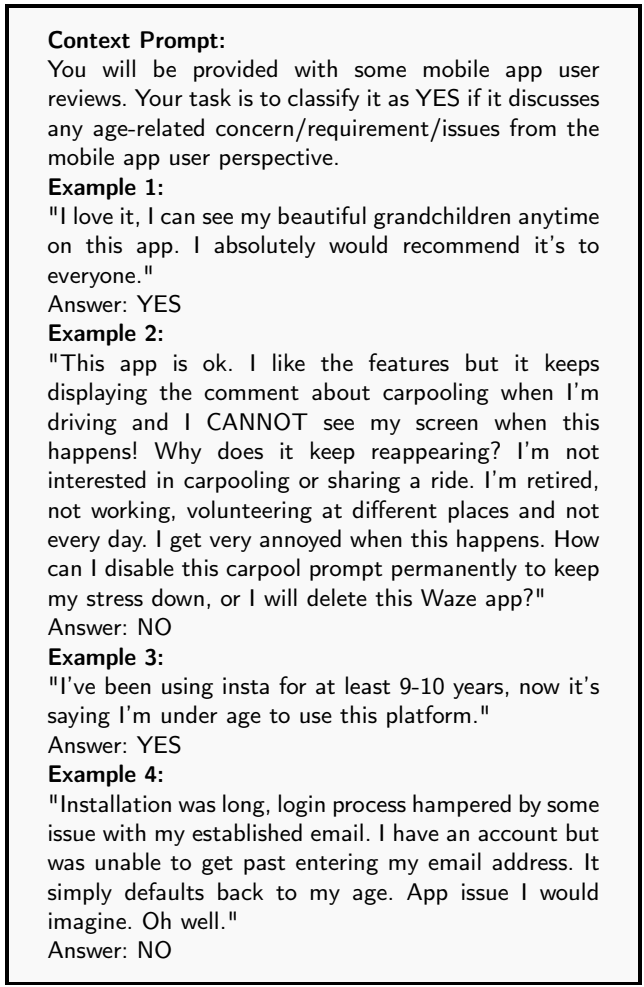

\begin{tcolorbox}[colframe=black, colback=gray!5, sharp corners]
\small
\textbf{Context Prompt:} \\
You will be provided with some mobile app user reviews. Your task is to classify it as YES if it discusses any age-related concern/requirement/issues from the mobile app user perspective.

\textbf{Example 1:} \\
"I love it, I can see my beautiful grandchildren anytime on this app. I absolutely would recommend it's to everyone." \\
Answer: YES

\textbf{Example 2:} \\
"This app is ok. I like the features but it keeps displaying the comment about carpooling when I'm driving and I CANNOT see my screen when this happens! Why does it keep reappearing? I'm not interested in carpooling or sharing a ride. I'm retired, not working, volunteering at different places and not every day. I get very annoyed when this happens. How can I disable this carpool prompt permanently to keep my stress down, or I will delete this Waze app?" \\
Answer: NO

\textbf{Example 3:} \\
"I've been using insta for at least 9-10 years, now it's saying I'm under age to use this platform." \\
Answer: YES

\textbf{Example 4:} \\
"Installation was long, login process hampered by some issue with my established email. I have an account but was unable to get past entering my email address. It simply defaults back to my age. App issue I would imagine. Oh well." \\
Answer: NO
\end{tcolorbox}

\captionsetup{justification=centering, font=small}
\caption{GPT Prompt with context} \label{GPT_prompt}
\end{figure}

As shown in Table \ref{tab:quant_results}, experiments were conducted under zero-shot and few-shot settings for each LLM. In the zero-shot setting, a prompt was designed to read the mobile app review and assign it a category without specifying enough context for what we mean by age-related discussion. In the few-shot setting, we designed a prompt with a detailed description of age-related discussion and a couple of examples from each case. Figure \ref{GPT_prompt} shows the few-shot GPT prompt. 

Two major concerns of LLMs are hallucinations and non-determinism. Non-determinism, in simple terms, is the problem of not always producing the same output for the same input \cite{ouyang2025empirical}. Hallucinations are defined as generated content that is either irrational or not faithful to the provided original content \cite{huang2025survey}. To mitigate these issues, setting \textit{temperature} and \textit{top P} parameters in the LLM models to near-zero values is recommended in the literature \cite{ouyang2025empirical,perkovic2024hallucinations}. Temperature is a parameter to control the randomness of the generated output \cite{Kimothi2023LLMParameters}. The \textit{top P} parameter controls the cumulative probability distribution that guides the selection of the next token during text generation \cite{Kimothi2023LLMParameters}. Hence, we set these values to 0.1 in our experiments to mitigate these issues.

The labels generated from the LLMs were then used along with the manually annotated labels to calculate the model's performance. The performance achieved by the LLMs in different scenarios was used to analyze the understanding of the LLMs regarding the age-related concepts in mobile app reviews.

\subsection{Experimental Settings} 
 \textcolor{black}{For BERT, RoBERTa, and DistilBERT models, input text was truncated to 256 tokens. \textcolor{black}{ Although RoBERTa supports input lengths of up to 512 tokens, we set the maximum input length to 256 to balance memory usage and training efficiency. This choice is supported by our data analysis, which showed that the median length of app reviews in our dataset is well below 256 tokens.} We set batch size to 16 for training ( 8 for validation )}. \textcolor{black}{For fine-tuning transformer-based models, small batch sizes (8–32) are commonly used to ensure stable gradient updates and generalization \cite{li2023codeeditor,zhang2024appt}. Accordingly, we selected a batch size of 16 for training. A smaller batch size of 8 was used during validation to reduce memory overhead, as backpropagation is not required.
 We experimented with multiple learning rates and found 1e-6 to yield the most stable and best-performing results.}

The ML and DL models mentioned above were trained using the PyTorch~\cite{paszke2019pytorch} library with a binary cross entropy loss function. A dropout layer was used after using the base model architecture, followed by the final classification layer. Except for the LLM-based experiments, the training was done using 10-fold cross-validation, where a model was trained for ten epochs in each fold. K-fold cross-validation (K=10 in this study) helps to achieve better model generalization even with a smaller dataset. This cross-validation technique is also helpful in preventing overfitting on the training dataset and hence provides more stable results. In this study, a test set was separated at each fold and later used to calculate the testing accuracy. The final performance was calculated as the average testing accuracy obtained in ten folds.

\textcolor{black}{
We evaluated GPT-4.1 (released on 14.04.2025), Gemini 2.0 Flash (released on 05.02.2025), and LLaMA-4-Scout-17B-16E-Instruct (released on 05.04.2025) in zero-shot and few-shot settings. Few-shot prompts included task descriptions and representative examples (Figure \ref{GPT_prompt}). To reduce non-determinism and hallucination, temperature and top-p values were set to 0.1. For the LLM-based experiments, performance was evaluated between ground truths (pre-annotated) and the labels predicted by the LLM.}

\textcolor{black}{
We report accuracy, precision, recall, and F1-score, following standard definitions. Scores reflect the ability to distinguish age-related from non-age-related reviews (binary classification). For ML and DL models, we report average performance across cross-validation folds. For LLMs, we provide single-pass results.
}

\subsection{Evaluation Metrics}
We used the standard ML/DL metrics (accuracy, precision, recall and F1 score) to evaluate the performance of different experiments in classification \cite{abdalkareem2020machine, yang2022survey}. These metrics rely on four different result types from the confusion matrix for the evaluation, which are True Positive (TP), False Positive (FP), True Negative (TN), and False Negative (FN). These four result types are used to define the evaluation matrices used in the study. 

Accuracy is used to signal the overall effectiveness of the learning models. It is the ratio of correct predictions to the total number of predictions. Precision focuses on the correctness of positive predictions and is the proportion of true positive predictions out of the total number of positive predictions. Furthermore, recall is used to estimate the completeness of positive predictions by calculating the ratio of the total number of positive predictions to the total number of actual positive samples. F1 score is the harmonic mean of precision and recall, used to balance the trade-off between precision and recall. 

\subsection{Quantitative Results}

\begin{table*}[]
\caption{Quantitative results of RQ1: The performance of the models}
\label{tab:quant_results}
\begin{tabular}{|l|c|c|c|c|}
\hline
Model                        & \multicolumn{1}{l|}{Accuracy} & \multicolumn{1}{l|}{Precision} & \multicolumn{1}{l|}{Recall} & \multicolumn{1}{l|}{F1 Score} \\ \hline
USE + SVM poly               & 89.33                         & 89.41                          & 89.33                       & 89.15                         \\ \hline
USE + XGBoost                & 87.41                         & 87.37                          & 87.41                       & 87.31                         \\ \hline
BERT                         & 91.71                         & 91.98                          & 91.71                       & 91.77                         \\ \hline
RoBERTa                      & \textbf{92.39}                & \textbf{92.70}                 & \textbf{92.39}              & \textbf{92.45}                \\ \hline
DistilBERT                   & 90.79                         & 91.07                          & 90.79                       & 90.87                         \\ \hline
GPT 4.1 (Zero Shot)          & 86.26                         & 75.01                          & 90.75                       & 82.14                         \\ \hline
GPT 4.1 (Few Shot)           & 87.10                         & 74.89                          & 94.69                       & 83.63                         \\ \hline
Gemini 2.0 Flash (Zero Shot) & 81.60                         & 67.89                          & 89.44                       & 77.19                         \\ \hline
Gemini 2.0 Flash (Few Shot)  & 84.05                         & 73.24                          & 85.37                       & 78.84                         \\ \hline
LLaMA 4-Scout (Zero Shot)    & 75.23                         & 60.76                          & 81.44                       & 69.60                         \\ \hline
LLaMA 4-Scout (Few Shot)     & 81.62                         & 69.54                          & 83.99                       & 76.09                         \\ \hline
\end{tabular}%
\end{table*}

Table \ref{tab:quant_results} shows the performance of different models in classifying mobile app reviews as age-related reviews. As the experiments were conducted using 10-fold cross-validation, the results shown in the table are average scores over the ten folds (except for LLMs).
RoBERTa achieved the highest accuracy, precision, recall and F1 score among all the classifiers. Other transformer-based models like BERT and DistilBERT also achieved comparable performance as RoBERTa with approximately a 1\% difference in classification accuracy.

Among other traditional ML models discussed previously, SVM performs better than XGBoost, with a slight difference in accuracy. SVM and XGBoost are popular ML models based on finding the optimal decision boundary and ensemble model, respectively. The high performance of SVM supports the clear boundary between the two classes.

For the LLM-based experiments, the experiment with few-shot settings performed better than the zero-shot settings for Gemini and LLaMA models. 
For all the ML and DL models, the performance achieved (in terms of accuracy, precision, recall and F1 score) is on par over the ten folds. The standard deviation of the performance scores for ten folds is less than one. This trend indicates the stable learning of the models for the classification.

\begin{figure}
\includegraphics[width = 110pt,height=90pt]{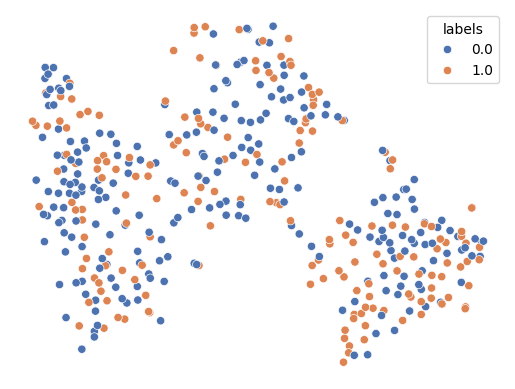}
\hspace{-5pt}
\includegraphics[width = 110pt,height=90pt]{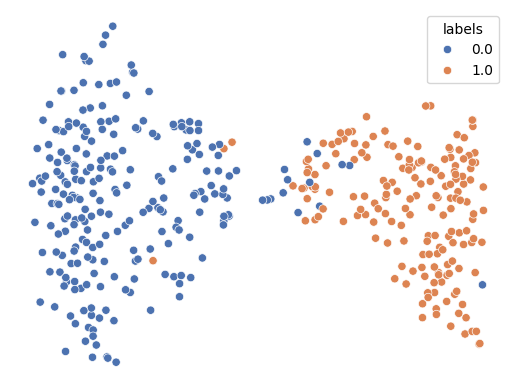}
\caption{t-SNE plot before (left) and after (right) training the best performing RoBERTa model. The plot is created using the test set for the fold corresponding to the best model.}
\label{fig:tsne_results}
\end{figure}

\subsection{Qualitative Results}

Figure~\ref{fig:tsne_results} shows the t-SNE~\cite{van2008visualizing} plots before and after training the RoBERTa model. t-SNE plot helps understand the high-dimensional features into 2 or 3 dimensions. Data points are mapped from the higher to lower dimensions so that their distance remains equivalent. For example, points closer to the high dimension also remain closer to the low dimension. To create the t-SNE plot, we used the best-performing RoBERTa model trained at the fifth fold. This model achieved around 94.7\% accuracy, precision and recall (with slight variation in the score). The test set corresponding to the fifth fold is used to create both t-SNE plots for comparable results.

The RoBERTa model is able to classify the two classes with clear class boundaries. For instance, the following two user reviews are both classified as age-related, and the features corresponding to these reviews achieved very high cosine similarity (0.989, where a distance closer to 1 indicates very high similarity). \emph{``I love this game. Lots of in game contests and chances to win boosters. Some levels, however, can be challenging. Great for all ages.''}, and \emph{``[Name] is awesome game it is for all ages but it's way better on xbox!''}. These reviews consist of mixed sentiments and thoughts. However, both suggest that the game is suitable for all age groups. The model is able to encode the fine details with the understanding of the overall task of classifying the age-related user reviews.

\section{Categories of Age Discussions (RQ2)} \label{rq2}

\subsection{Approach}

\begin{figure*}[t]
  \centering
  \includegraphics[width=0.90\linewidth]{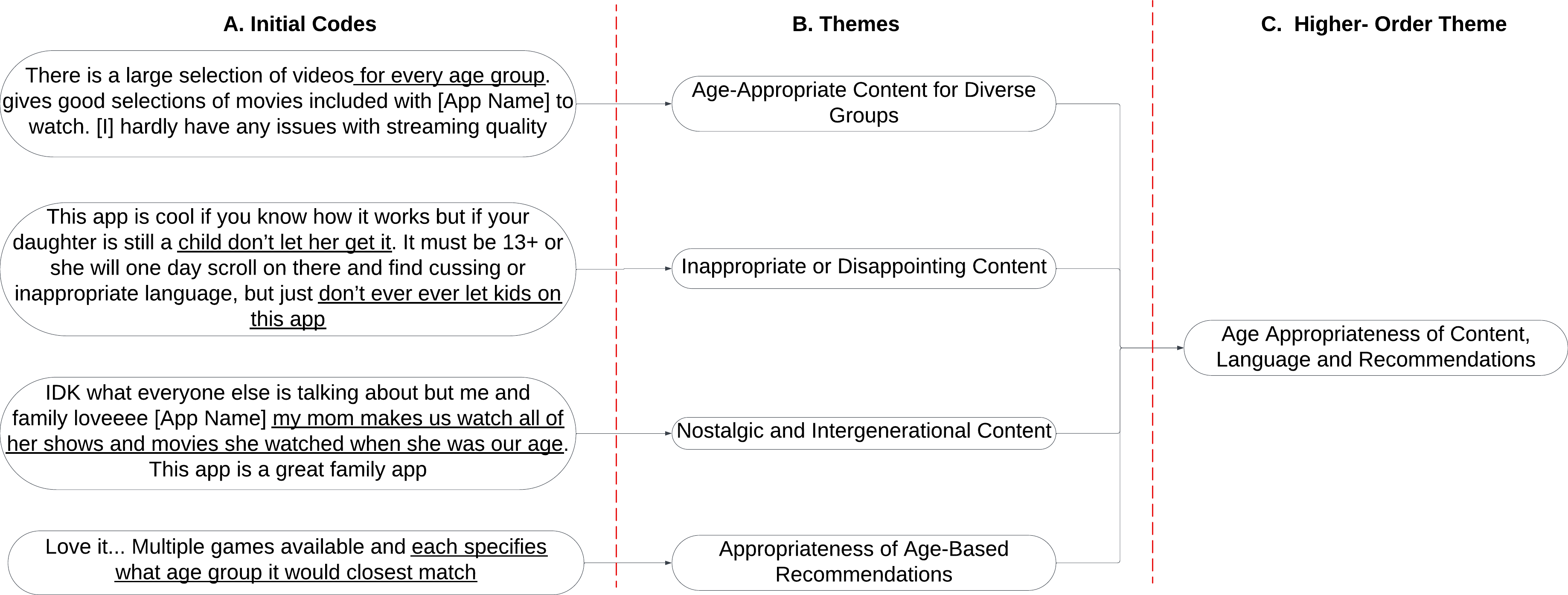}
  \caption{\textcolor{black}{The steps of performing thematic analysis.}}
  \label{fig:thematic-flow}
\end{figure*}

To understand the types of age discussions in the mobile app reviews, thematic analysis was performed by the first author on 1,429 previously annotated age-related user reviews (detailed in Section \ref{sec_dataset}).
\textcolor{black}{We followed established guidelines for thematic analysis and adapted procedures previously used in SE research \cite{braun2006using}. The analysis involved multiple iterative stages:
\begin{enumerate}
    \item Familiarizing with data: The first author read all age-related reviews multiple times to become deeply familiar with the content and to obtain an initial sense of age-related issues expressed by users.
    \item Generating initial codes: The first author systematically examined each review and assigned descriptive codes to meaningful text segments without using a predefined coding scheme (Fig. \ref{fig:thematic-flow}. A).
    \item Searching for themes: Related initial codes were then grouped together to form potential themes and sub-themes that represented recurring patterns across the dataset (Fig. \ref{fig:thematic-flow}. B).
    \item Reviewing themes: The preliminary themes and their associated reviews were independently examined by the third and fourth authors. Multiple review meetings were held among the three analysts to discuss the themes and resolve potential conflicts using the negotiated agreement approach \cite{campbell2013coding}. Ultimately, a common consensus among all three analysts was generated regarding the final themes and higher-order themes.
    \item Defining and naming themes: Finally, clear and meaningful names were assigned to each theme and higher-order theme to accurately represent the age-related concerns identified in user reviews. This process resulted in six higher-order themes and seventeen sub-themes that captured how users discuss age in mobile app reviews (Fig. \ref{fig:thematic-flow}. C).
\end{enumerate}
}

\textcolor{black}{This approach is widely adopted in qualitative and mixed-methods research within the software engineering community \cite{nasab2024study, khalajzadeh2022supporting, chattopadhyay2020tale}, particularly when the goal is iterative theme development and conceptual refinement rather than independent coding comparison.}
\textcolor{black}{Fig \ref{fig:thematic-flow} shows the application of thematic analysis to analyze various reviews and identify the ‘‘Age Appropriateness of Content, Language and Recommendations’’ theme.}

\subsection{Results}
This section elaborates on the 6 higher-order themes and 17 themes of age discussions we identified in app reviews (see table \ref{themes}). We use the notation $\times$\emph{\textbf{i}} to indicate that \emph{\textbf{i}} reviews mentioned the corresponding theme.

\begin{table*}[]
\caption{Higher-order themes and themes of age discussions in mobile app user reviews.}
\label{themes}
\begin{tabular}{|l|l|}
\hline
Higher Order Themes & Themes \\ \hline
\multirow{4}{*}{\begin{tabular}[c]{@{}l@{}}Age Appropriateness of Content,\\ Language and Recommendations (925)\end{tabular}} & Age-Appropriate Content for Diverse Groups (664)\\ \cline{2-2} 
 & Inappropriate or Disappointing Content (215) \\ \cline{2-2} 
 & Nostalgic and Intergenerational Content (33) \\ \cline{2-2} 
 & Appropriateness of Age-Based Recommendations (13) \\ \hline
\multirow{4}{*}{Age Verification and Access Barriers (248)} & Technical Verification and Restriction Issues (132) \\ \cline{2-2} 
 & User Control Over Age Settings (52) \\ \cline{2-2} 
 & Parental Control Gaps (17) \\ \cline{2-2} 
 & Overly Strict or Unnecessary Limits (47)\\ \hline
 \multirow{2}{*}{\begin{tabular}[c]{@{}l@{}}Usability and Accessibility \\ Across Ages (109)\end{tabular}} & Challenges for Specific Age Groups (56) \\ \cline{2-2} 
 & Positive User Experience Across Ages (53) \\ \hline
\multirow{2}{*}{Privacy and Safety Concerns (95)} & Risks for Minors (80) \\ \cline{2-2} 
 & Unnecessary Personal Data Collection (15)\\ \hline
\multirow{2}{*}{Interactions and Relationships (55)} & Unsafe Interactions(30) \\ \cline{2-2} 
 & Positive Relationships (25) \\ \hline
\multirow{3}{*}{\begin{tabular}[c]{@{}l@{}}Recommendations and Feature \\ Requests (52)\end{tabular}} & Age-Specific Feature Demands (18)\\ \cline{2-2} 
 & Policy or Verification Changes (15)\\ \cline{2-2} 
 & Content Improvements (19)\\ \hline
\end{tabular}%
\end{table*}

\subsubsection{Age Appropriateness of Content, Language and Recommendations ($\times$925)} 

Most of the user reviews in this category appreciated well-curated content that aligns with the age expectations of kids, teenagers, and the elderly. They also acknowledge that the content is suitable for all ages. For example,
\newline
\faThumbsOUp \emph{“There is a large selection of videos for every age group. gives good selections of movies included with [App Name] to watch. [I] hardly have any issues with streaming quality.”}

However, many reviews mention dissatisfaction with inappropriate content for certain age groups, particularly children and teenagers. Further, this includes concerns about inappropriate language and the lack of engaging content for kids and adults.
\newline
\faThumbsODown \emph{“I think [App Name] is played even by some children and youngsters and to win boosters, sometimes you are asked to watch advertisement or video which shows inappropriate and obscene videos which are not to be watched by children or teenagers. Dear parents, If your children are playing be careful.”}
\newline
\faThumbsODown \emph{“This app is cool if you know how it works but if your daughter is still a child don't let her get it. It must be 13+ or she will one day scroll on there and find cussing or inappropriate language, but just don't ever ever let kids on this app”}

Some users appreciate apps that offer nostalgic content, particularly those that allow parents to share their cherished childhood experiences with their children.
\newline
\faThumbsOUp \emph{“IDK what everyone else is talking about but me and family loveeee [App Name] my mom makes us watch all of her shows and movies she watched when she was our age. This app is a great family app”}

Some users appreciate the age-appropriateness of recommendations, while others express dissatisfaction when recommendations do not align with their age. For example,
\newline
\faThumbsOUp \emph{“Love it... Multiple games available and each specifies what age group it would closest match.”}
\newline
\faThumbsODown \emph{“[App Name] is giving me notifications on my device that are very inappropriate for my age group….”}

\subsubsection{Age Verification and Access Barriers ($\times$248)} 

Users frequently report issues with age verification processes, such as incorrect calculations leading to account restrictions or age filters failing to function properly.
\newline
\faThumbsODown \emph{“My account is locked only because this application thinks I haven't reached the age of 13, I just filled in the date, month, and year of my birth, and suddenly my account get locked. Even though my age has exceeded 13. Please repair it, before I uninstall it, bad.”}
\newline
\faThumbsODown \emph{“Too much idiocy and depraved trash for kids. The filters and restricted mode don't seem to work...”}

Some reviews mention frustration over being forced to verify their age, sudden restrictions, or the inability to change their age settings after an initial input error. For example,
\newline
\faThumbsODown \emph{“There's a reason that there are so many 3rd party apps. Such a terrible design to force people to open in the app to “verify age” on text posts. Horrible site, Horrible app, Horrible company.”}
\newline
\faThumbsODown \emph{“I had the app for a year and it is just now banning my account because I'm not up to age.”}
\newline
\faThumbsODown \emph{“When you accidentally put in the wrong age, it won't let you go back to change it.”}

While some users appreciate the availability of parental controls, many express concerns that the existing controls are inadequate or missing key features. For example,
\newline
\faThumbsOUp \emph{“Great internet tools, allows for a ton of privacy control and easy to use features too keep content out of children's hands.”}
\newline
\faThumbsODown \emph{“The parental controls for [App Name] are a joke. You can't block specific shows, and the age ratings are not always appropriate. Considering canceling my membership because of the limited parental controls. How hard would it be for them to add a block button to the stupid shows they include on [App Name]?”}

Users complain about restrictions that feel excessive, particularly for older users. Further, teenagers or kids near the age limit often express their disappointment.
\newline
\faThumbsODown \emph{“I love the app, but a lot of videos get either unnecessarily demonized or age limited, especially political content...”}
\newline
\faThumbsODown \emph{“[App Name] is fun and their is alot of creativity. But the [App Name] team should really take down the no messaging people under the age of 16. a lot of creaters won't get as many shares as they used to because of this update...”}

\subsubsection{Usability and Accessibility Across Ages ($\times$109)}

Kids and elderly users express frustration with user interfaces that are difficult to navigate, have small text, and complex layouts.
\newline
\faThumbsODown \emph{“Does what it says. Only problem is when it changes (interface's text usually just gets smaller and harder to read) with each update, as the 'user experience' they claim to improve has steadily declined the usefulness of this app. It seems (like most apps) to be based around the physical attributes of people under the age of 45. The interface is only usable because of memory; I can't personally read the interface text anymore. Gets worse each update.”}

Some reviews praise apps for being user-friendly and easy to learn, especially for elderly users and kids.
\newline
\faThumbsOUp \emph{“I believe it to be very user-friendly, especially for people in my age bracket. I am 53 and have been learning without asking friends or family members, just by using this app. Keep up the good work.”}
\newline
\faThumbsOUp \emph{“Even my six year old son uses Photos so easily. Thumbs up [App Name]!”}

\subsubsection{Privacy and Safety Concerns ($\times$95)} 

Users raise concerns about inappropriate content involving minors and children misrepresenting their age to gain access to apps.
\newline
\faThumbsODown \emph{“Reels often include inappropriate videos of very young kids. Reporting the posts only generates more of those types of videos...”}
\newline
\faThumbsODown \emph{“A lot of kids are using [App Name] inappropriately. I've seen a lot myself, they lie about their age and put an age appropriate for their age, even though they're way too young. I've been using this app for 4 years now and it's the best, just about the young kids lying about their age and posting inappropriate stuff.”}

Some users express concerns about how apps handle personal data, particularly regarding the information collected from minors.
\newline
\faThumbsODown \emph{“Absolutely the most overhyped app... It only does damage to kids. It has no other purpose than to do that and to collect private data. FOR YOUR OWN GOOD: DELETE IMMEDIATELY!”}

\subsubsection{Interactions and Relationships ($\times$55)}
Some users worry about unwanted or unsafe interactions, particularly for minors. Further, users express concerns about deceptive interactions like scams targeting specific age groups. 
\newline
\faThumbsODown \emph{“I have a lot of minors trying to message me and that is NOT ok because that means it's just as easy for an adult to message them! Step up the security [App Name], these are our children, our future, we need to protect them better. (In my opinion)”}
\newline
\faThumbsODown \emph{“Pretty well designed game, I would give it a 5, but there are setbacks such as lots of scammers taking advantage of the younger age group…”}

Users highlight the benefits of apps in maintaining family relationships or fostering social interaction across different age groups.
\newline
\faThumbsOUp \emph{“Use it daily to stay in touch with my kids and do occasionally hear from my grand kids !...”}
\newline
\faThumbsOUp \emph{“Best social media app that connects people of any race, age and the world at large”}

\subsubsection{Recommendations and Feature Requests ($\times$52)} 

Reviewers suggest introducing versions of apps tailored for kids, enabling users to create multiple profiles to prevent mixing content with family members of different ages, and allowing users to set age-based interaction limits.
\newline
 \faThumbsODown \emph{“I wish people under 18 could get a card, kids need money too, maybe you could make a [App Name] app just for kids so the parents can send their kids money”}
\newline
\faThumbsODown \emph{“It would be great if you could implement some type of basic profiles, like [App Name], [App Name] and other streaming services, so I can avoid polluting my profile with kids songs and other songs I don't like...”}
\newline
\faThumbsODown \emph{“They should let us put an age limit for people to watch because old men watch my videos and it scares people.”}

Users propose changes to age-related policies, such as removing certain age restrictions or allowing payment for underage individuals.
\newline
\faThumbsODown \emph{“Need to remove age limit for parents to teach their kids about proper money spending”}

Some users request more family-friendly content or content in additional languages to cater to diverse audiences.
\newline
\faThumbsODown \emph{“Needs more state languages, good family and kid movies...”}

\section{Discussion} \label{discussion}

\textbf{\textit{Enhancing User Experience by Rethinking Rigid Age Restrictions.}} Our analysis indicated a significant dissatisfaction with strict age-based restrictions, impacting both adolescents and older adults facing unnecessary limitations. A recent study emphasized the negative effects of rigid, all-or-nothing parental control systems for adolescents \cite{wang2021protection}. These systems often lead parents to impose overly restrictive settings and intrusive surveillance. Participants in the study expressed significant frustration and distress over their loss of autonomy and privacy, which led to mental health issues \cite{wang2021protection}. Additionally, the findings indicated that these strict restrictions prevented children from accessing basic, harmless, and essential features of their devices that they deemed important \cite{wang2021protection}.
Conversely, older users complained of patronizing content filters and feature locks. Moreover, a recent study revealed that social media apps often misapply restrictions, mistakenly flagging or removing content under ``Minor Safety" policies \cite{zeng2022content}. Further, they revealed that these apps fail to effectively prevent actual underage access, which discourages older content creators.

\textcolor{black}{From a technical perspective, many of these frustrations stem from rule-based access controls and fixed policy enforcement pipelines that do not adapt to nuanced edge cases \cite{yaoeasy,patnaik2021don}. Incorporating more dynamic permission logic and configurability could help accommodate borderline age scenarios without compromising safety.}

\textbf{\underline{Implications.}}
Developers should consider implementing more flexible or gradient-based age restrictions that account for borderline cases. For example, instead of a hard cutoff at age 13, apps could introduce tiered access or parental approval for features when users are within the age limit.  For younger children, these apps could start with the highest level of restrictions and monitoring, gradually easing these measures as the child grows older or demonstrates the ability to identify and manage online risks \cite{wang2021protection}. Rather than relying solely on platform-wide moderation systems that often frustrate users through overblocking or inconsistent enforcement, researchers are advocating for personal content moderation as an alternative approach \cite{jhaver2023personalizing}. It enables users to customize their moderation experience by adjusting filtering preferences and setting their own thresholds for acceptable content based on the content submitted by other users  \cite{jhaver2023personalizing}. 

\medskip
\textbf{\textit{Tackling Age Falsification and Strengthening Age Verification Mechanisms.}} 
Issues related to age verification were discussed in 248 reviews analyzed, making it the second main theme in age-related discussions. A recurring concern in the reviews was minors bypassing age restrictions by falsifying their age during account creation. This loophole exposes younger users to inappropriate content or unsafe interactions, undermining the app’s safety measures. Current age verification methods are divided into two categories: offline and online. Offline methods, such as age gates and ID checks against databases, are simple but less reliable \cite{yaoeasy}. In contrast, online methods, which include ID validation, credit card checks, document uploads with OCR, and biometric analysis, are more accurate and scalable\cite{yaoeasy}. However, online methods also raise greater privacy concerns and involve higher implementation costs compared to offline approaches.

Moreover, users also reported technical glitches in age verification systems, such as incorrect age calculations or irreversible errors after initial input (e.g., locking accounts erroneously). These issues highlight the inadequacy of current verification methods, which often rely on self-reported data or simplistic checks.
\textcolor{black}{These failures are often linked to brittle input-validation logic and limited exception-handling pathways, where account-creation workflows do not provide rollback mechanisms or safe correction flows \cite{yaoeasy,patnaik2021don}.}

\textbf{\underline{Implications.}}
App developers can improve age verification by using strategies such as Multi-Factor Verification (MFV), which combines ID verification with biometric authentication and checks ID details against official government databases in real time \cite{yaoeasy}. Moreover, with the current advancements in AI, researchers may investigate the potential for using AI in age verification. AI-driven age estimation using facial images has demonstrated a high degree of accuracy in determining a person’s age \cite{rahman2023attention, bekhouche2024facial, george2024chronological}. For example, a recent study focused on an AI-powered age verification system that utilizes low-quality facial images \cite{wargo2024privacy}. The findings showed that this system successfully balances accurate age verification with the protection of user privacy, as it minimizes the need for intrusive data collection.

\medskip
\textbf{\textit{Personalization of Content Based on Age Groups.}}  According to several reviews, users praised the versatility and family-friendly nature of certain apps that allowed users of all ages to enjoy the app together. However, our study reveals that users raised concerns regarding the availability of inappropriate content for children, noting that some content presented in mobile apps is unsuitable for specific age groups and could be made more age-friendly. Studies \cite{chen2013children} \cite{zhao2023mobile} found that a significant portion of in-app ads feature content that is unsuitable for children. Further, Meyer et al.\cite{meyer2019advertising} revealed that in-app purchase facilities or offers to upgrade to a full version were promoted in unethical ways for young children in free apps. For example, in-app purchases are often advertised through familiar characters in the app, which children are likely to trust and feel emotionally connected to. This strategy leverages children's parasocial relationships with these characters to encourage spending, which can be considered a misuse of children's trust and emotional bonds with media figures. In addition, Zhao et al. \cite{zhao2023mobile} revealed that certified ad SDKs could still distribute unsuitable adverts. To improve the ad libraries, Ahasanuzzaman et al. \cite{ahasanuzzaman2020studying} suggested that ad library developers should improve external ad-mediators by enabling integration of new ad libraries at run-time, increasing supported ad libraries, and providing more feature control over ad library selection.

\textcolor{black}{Technically, implementing such personalization requires modular content-delivery pipelines, metadata tagging for age relevance, and ranking mechanisms that respond to user profiles and safety constraints. Poorly tuned ad-delivery SDKs and content classifiers can lead to misalignment between expected and delivered content for different age groups.\cite{alomar2022developers,carlsson2023privacy}}

\textbf{\underline{Implications.}}
We argue that users of different age groups have varying needs, expectations, and sensitivities, which should be considered when curating content. There is a need for more advanced algorithms that personalize content based on the user's age-related preferences, along with associated generational cultural and lifestyle preferences. Further, filters can be introduced to automatically adjust content based on the user's age group. It is essential to ensure content personalization extends not just to the app's core features but also to its subordinate features, like advertisements. Future research can explore the ethical concerns of the current app content and improve ad SDKs. 

\medskip
\textbf{\textit{Enhancing Digital Safety and Well-being for Young Mobile Apps Users.}} As digital natives, children are increasingly interacting with mobile apps that may not be adequately designed with their safety and well-being in mind. Different types of child safety concerns were raised in several app review categories. For example, inappropriate content for children, the interaction of children with strangers, the need for improved parental control, exposing children to bad behaviour and learning, etc. Several studies have been conducted to understand child safety concerns in mobile apps. Their findings revealed that 96\% of the most popular free apps on Google Play for kids under five had commercial content, such as interstitial advertising and hidden ads that pop up automatically \cite{radesky2020digital}. Moreover, STEM apps in Google Play and Apple App Store for young children may not be developmentally appropriate, contain distracting features, or have limited educational content \cite{konca2024evaluating}. The past work revealed that parental controls in mobile apps mainly focus on immediate threats like cyberbullying or inappropriate content \cite{zhao2022koala}. They offer limited assistance to children in understanding the dangers of online data privacy and the harms that come with it. Further, Papadakis et al. \cite{papadakis2019parental} found that Greek parents often lack the knowledge and experience to effectively evaluate educational apps. As a result, they may struggle to incorporate these apps appropriately into the home environment to educate their children. Another study raised concerns about low-quality content and distracting visual and audio elements in popular children's apps. Instead of using unnecessary visual or audio effects, they emphasized the need for developing interactive components that assist learning objectives \cite{hirsh2015putting}. 

\textbf{\underline{Implications.}} 
Developers need to prioritize safety features in mobile apps designed for children. This includes filtering out inappropriate content, providing robust parental controls, and ensuring that no interactions with strangers are allowed. To filter inappropriate content, moderation systems should be implemented to automatically detect and remove any content with offensive language, violence, or adult themes. Parental controls should allow them to customize the app experience by setting limits on screen time, restricting access to certain content, and monitoring app usage. Furthermore, apps must include secure communication features where only approved contacts can communicate with the child. Another important step is educating children, which is crucial for ensuring data protection while using mobile apps \cite{senigaglia2022social}. Further, we highlight the necessity of increasing parental awareness and understanding of app functionalities to create a safer digital environment for young mobile app users.

\medskip
\textbf{\textit{Extending Age-Inclusive Analysis Beyond App Reviews.}} 
While our ML, DL, and LLM models (e.g., RoBERTa) can automatically detect age-related app reviews with high precision (92.70\%), we relied on the manual thematic analysis to categorize the types of age discussions (e.g., “Age Verification and Access Barriers,” “Privacy and Safety Concerns”). Although this method ensured rigor and depth, it constrained the dataset size due to the time-intensive nature of manual annotation. The LLMs we evaluated do not require labelled training data and showed promising capabilities in identifying age-related reviews. Although these LLMs did not outperform RoBERTa, their generalization ability and minimal setup requirements make them attractive tools for extending age-related analysis to broader and less-structured data sources.

Moreover, our current focus on app reviews captures only a narrow slice of the software development lifecycle. Age-related concerns may also arise in other artifacts, such as GitHub issues (e.g., accessibility feature requests), UX design documents, or automated test cases. For instance, a study was conducted on Reddit\footnote{https://www.reddit.com/} to explore public perceptions regarding age in the software development industry \cite{serebrenikage}. These artifacts can surface unique challenges related to technical feasibility, design intent, and policy enforcement, all of which are crucial to creating age-inclusive software. As LLM capabilities continue to improve, they may serve as key enablers in scaling age-inclusivity analysis to these additional software artifacts without requiring annotated training data.

\textbf{\underline{Implications.}} To extend age-inclusivity analysis across the software lifecycle, researchers should develop automated methods that can classify and summarize age-related content across diverse artifacts. For example, models that jointly analyze sentiment and age-related context can help prioritize age-relevant issues that also express user frustration, enabling better triaging (e.g., older users struggling with complex navigation). Additionally, discourse analysis tools can be used to compare how developers frame age-related issues (e.g., in GitHub discussions) with how end-users report them (e.g., in reviews), revealing potential misalignments in priorities or understanding \cite{walters1992discourse}. These methods will enable more comprehensive and continuous age-inclusivity monitoring across the entire development pipeline. Future research should explore how emerging LLMs can support a nuanced understanding of age-related discussions across multiple domains, ultimately reducing reliance on manual labelling.

\section{Threats to Validity} \label{threats}
This section discusses the threats to the validity of our study and the strategies we adopted to mitigate the threats.

\textbf{Internal Validity }
The process of curating 29 n-grams for the taxonomy was based on existing literature and general understanding. The list of n-grams might be limited to include every possible age-related discussion. However, the chosen n-grams offer a sufficient representation of the age-related concepts found in the user reviews. A set of 29 n-grams is consistent with existing studies, which often use a similar number of terms to cover relevant concepts \cite{ebrahimi2022unsupervised}. Analyst bias may affect categorizing reviews as age or non-age-related (RQ1), and analysis of age reviews to determine the types of age discussions (RQ2).  To mitigate this threat, the processes were carried out in several iterations to prevent analyst fatigue and foster common understanding. Further, we ensured all analysts agreed on the final age discussion types in RQ2.

\textbf{Construct Validity}  We only employed 8 ML, DL and LLM models. Although these models are commonly employed in research and practice domains, we acknowledge that this represents only a subset of the many available models. Moreover, alternative models may yield different results \cite{peters2017text,shahin2023study}. Furthermore, even if we fine-tune the models' parameters, changing them may produce different outcomes.  The ML/DL models used in this study are specifically trained on a dataset of mobile app reviews. App reviews tend to be written in informal, conversational language, which may not translate well to more formal or structured texts that discuss age-related issues differently. Hence, these models may struggle to perform well when used on other types of data, such as product reviews from e-commerce sites or discussion threads on various forums, without retraining them on the new data. To validate the results of RQ2, additional research methods, such as surveying app users, could be employed in addition to mining reviews. 

\textbf{External Validity} Our dataset consists of user-generated app reviews, which may not represent the entire user population equally. Research suggests that certain demographic groups (e.g., children, elders) may be less likely to leave reviews compared to others \cite{pew2016online,picariello2016review}, potentially skewing the data toward specific age groups or user types. While we lack explicit demographic metadata to verify this bias, it remains a limitation that could affect the generalizability of our findings.
Moreover, the dataset created from the top 70 popular Android apps in the Google Play Store was used in this study \cite{shahin2023study}. It only contained a portion of the reviews from those 70 apps. Further, the dataset is in English and might come from users in specific geographic regions. This may not generalize to non-English speakers or users from different cultural backgrounds who might express age-related concerns differently. Lastly, we only analyzed 1,429 age-related app reviews to answer RQ2. This indicates the age discussions identified in RQ2 may not be generalizable and comprehensive for all apps in the Google Play Store, Apple App Store, or any other type of software system. Nonetheless, we argue that such threats were mitigated as the Android apps used in this analysis span a range of domains and categories and accumulate a substantial number of reviews that may indicate a certain level of representativeness. \textcolor{black}{Additionally, this dataset was collected in 2023 and has not been updated since its release. We acknowledge that the rapid evolution of mobile applications and the introduction of new AI-driven features and interaction patterns may create age-related concerns that this dataset does not capture.  However, it represents the most comprehensive and recent publicly available dataset of this scale for age-related analysis in mobile app reviews.}

\textbf{Reliability} The replication package for this study is available online, which can be used for future research \cite{anonymous2025age}. It includes the dataset with all annotations, source codes for ML and DL models, LLM prompts, and results.

\section{Conclusion and Future Work} \label{conclusion}

In this study, we explored age discussions in user reviews using 70 popular Android apps in the Google Play Store. We aimed to understand user discussions related to age, in order to help develop age-inclusive apps. We developed eight ML and DL classifiers to automatically detect app reviews containing age discussions. These classifiers were trained and evaluated on a manually curated dataset of 4,163 app reviews, which contained 1,429 age-related reviews and 2,734 non-age-related reviews. The RoBERTa model performed the best in categorizing reviews as age-related and non-age-related. It achieved precision, recall, F1-score, and accuracy of 92.70\%, 92.39\%, 92.45\%, and 92.39\%, respectively. Moreover, a qualitative analysis of 1,429 age-related reviews provided further insights into six major topics of age discussions and a set of different concepts discussed under each main topic. These topics captured both the aspects users appreciated about the apps and the challenges they encountered. This analysis highlights the areas where the apps succeeded and where improvements were needed to cater to specific age-related concerns. 

For future work, we will expand our research by examining age discussions in other software domains, such as Q\&A platforms like Stackoverflow. Furthermore, we will investigate how age discussions vary across different app domains, such as health, education, social media, and entertainment, to understand how app categories influence user perspectives on age inclusivity. Finally, we aim to enrich our understanding of age discussions by employing mixed-method approaches, such as conducting surveys or interviews with app users to validate the qualitative insights gained from review mining.

\section{Acknowledgments}
This project is supported by Deakin University's Career Continuity for Researchers who are Primary Carers (CCRPC) Fund.

\bibliographystyle{unsrtnat}  
\bibliography{age_main}

\end{document}